\documentclass[12pt,a4paper]{article}

\usepackage{amsmath,amsfonts,latexsym,amssymb,amscd}
\usepackage{pslatex}
\usepackage[latin1]{inputenc}
\usepackage[T1]{fontenc}
\usepackage{verbatim,amsthm,curves,graphics}
\usepackage{mathrsfs}

\def\ben{\begin{equation}}

\def\een{\end{equation}}

\def\bena{\begin{eqnarray}}

\def\eena{\end{eqnarray}}

\def\f(#1/#2){\frac{#1}{#2}}

\def\Frac(#1/#2){\left(\frac{#1}{#2}\right)}

\def\chris(#1-#2-#3){{\mit \Gamma}^{#1}{}_{{#2}{#3}} }

\def\tilchris(#1-#2-#3){\tilde{{\mit \Gamma}}^{#1}{}_{{#2}{#3}}}

\def\hatchris(#1-#2-#3){\hat{{\mit \Gamma}}^{#1}{}_{{#2}{#3}}}

\newfont{\schwell}{schwell scaled 1101}

\newcommand{\non}{\nonumber}

\theoremstyle{definition}

\renewcommand{\epsilon}{\varepsilon}
\newcommand{\myid}{{\bf 1}}

\newcommand{\mc}{{\mathbb C}}

\newcommand{\M}{{\bf M}}
\newcommand{\C}{{\mathcal C}}

\newcommand{\A}{{\mathcal A}}
\renewcommand{\S}{{\mathcal S}}
\begin{document}

\title{Quantum field theory in curved spacetime, the operator product
  expansion, and dark energy}

\author{Stefan Hollands$^a$\thanks{HollandsS@Cardiff.ac.uk} \,, Robert
  M. Wald$^b$\thanks{rmwa@uchicago.edu} \\
  \\
  $^a$ {\em School of Mathematics, Cardiff University, UK} \\
  \\
  $^b$ {\em Enrico Fermi Institute and Department of Physics} \\
{\em University of Chicago, Chicago, IL, USA}
  \\
  \\
  \\}

\maketitle

\abstract{To make sense of quantum field theory in an arbitrary
  (globally hyperbolic) curved spacetime, the theory must be
  formulated in a local and covariant manner in terms of locally
  measureable field observables. Since a generic curved spacetime does
  not possess symmetries or a unique notion of a vacuum state, the
  theory also must be formulated in a manner that does not require
  symmetries or a preferred notion of a ``vacuum state'' and
  ``particles''. We propose such a formulation of quantum field
  theory, wherein the operator product expansion (OPE) of the quantum
  fields is elevated to a fundamental status, and the quantum field
  theory is viewed as being defined by its OPE. Since the OPE
  coefficients may be better behaved than any quantities having to do
  with states, we suggest that it may be possible to perturbatively
  construct the OPE coefficients---and, thus, the quantum field
  theory. By contrast, ground/vacuum states---in spacetimes, such as
  Minkowski spacetime, where they may be defined---cannot vary
  analytically with the parameters of the theory. We argue that this
  implies that composite fields may acquire nonvanishing vacuum
  state expectation values due to nonperturbative effects. We
  speculate that this could account for the existence of a
  nonvanishing vacuum expectation value of the stress-energy tensor of
  a quantum field occurring at a scale much smaller than the natural
  scales of the theory.}

\section*{Quantum field theory in curved spacetime, the operator product
  expansion, and dark energy}

Quantum field theory in curved spacetime is a theory wherein matter is
treated fully in accord with the principles of quantum field theory,
but gravity is treated classically in accord with general relativity.
Despite its classical treatment of gravity, quantum field theory in
curved spacetime has provided us with some of the deepest insights we
presently have into the nature of quantum gravity. The main purpose of
this essay is to argue that it also is providing us with significant
insights into the nature of quantum field theory itself.

One of the key insights that has been obtained by the study of free
(i.e., non-self-interacting) quantum fields in curved spacetime is
that---apart from stationary spacetimes or spacetimes with other very
special properties---there is no unique, natural notion of a ``vacuum
state'' or of ``particles''. Indeed, unless the spacetime is
asymptotically stationary at early or late times, there will not, in
general, even be an asymptotic notion of particle
states. Consequently, it is essential that quantum field theory in
curved spacetime be formulated in terms of the local field observables
as opposed, e.g., to S-matrices. ``Particle detectors'' should be
viewed as systems that interact with the local quantum fields
\cite{unruh}, whose response can be calculated independently of any
notion of ``particles'' that one may wish to introduce \cite{uw}.

The usual formulations of quantum field theory in Minkowski spacetime
(of both the rigorous and pragmantic kinds) rely on the following
three key ingredients that do not generalize in an obvious way to
curved spacetimes: (i) Poincare covariance of the quantum fields; (ii)
positivity of total energy (spectrum condition); and (iii) the existence
of a Poincare invariant state (``the vacuum'').
With regard to (i), a generic curved
spacetime will not possess any symmetries at all, so no spacetime
symmetry requirements of any kind can be imposed on quantum fields in
generic curved spacetimes. With regard to (ii), in the absence of a time
translation symmetry, the total energy of even a classical field is
highly time-slice dependent; since the energy density of a quantum field
(in flat or curved spacetime) can be negative, and, in some simple
examples involving free fields in curved spacetime, the integrated
energy density is found to be negative, it seems clear that
in curved spacetime no useful
spectrum condition can be formulated in terms of the ``total
energy-momentum'' of the quantum field. Finally, with regard to (iii),
not only is ``Poincare invariance'' meaningless in curved spacetime,
but one cannot expect that there exists a sensible criterion for
picking out a special state of any kind in an arbitrary curved spacetime.

Nevertheless, in quantum field theory in curved spacetime, it is by
now understood that there does
exist a suitable replacement for the requirement, (i), of Poincare
covariance of the quantum fields. Rather than require that the quantum
fields be ``specially covariant'' in the sense of having the Poincare
group act as a symmetry group, we require the quantum fields to be
``generally covariant'' in the sense that they be locally and
covariantly constructed from the spacetime metric (as well as from other
``background structure'' such as space and time orientations and spin
structure). In order to define this notion, it is essential that the
quantum field theory be defined on {\it all} (globally hyperbolic)
curved spacetimes, since in order to express the idea that the fields
are locally determined by the spacetime metric, it is necessary to see
how the theory changes when we change the metric in an arbitrary way.
We then consider the following situation: Let $(M,g)$ and $(M',g')$ be
two globally hyberbolic spacetimes that have the property that there
exists an isometric imbedding, $\rho$, of $M$ into $M'$, which also
preserves causality relations and all of the other background
structure. We require that there be a corresponding isomorphism,
$\chi_\rho$, of the algebra of field observables on $(M, g)$ with the
subalgebra of field observables on $(M',g')$ associated with the
region $\rho[M]$. Furthermore, we require that $\chi_\rho$ maps each
smeared quantum field $\phi^{(i)}(f)$ on $M$ to the corresponding
quantum quantum field $\phi^{(i)}(\rho_* (f))$ on $M'$;
see~\cite{hw1,hw2,bfv} for further discussion.

In addition, there exists a suitable generalization of the spectrum
condition, (ii), to curved spacetime. The key idea is that the
``positive energy'' properties of quantum fields in Minkowski
spacetime are directly related to their ``positive frequency''
properties, which, in turn, are directly related to the short-distance
singularity structure of the of the $n$-point functions of the quantum
fields. The positive frequency nature of the short-distance
singularities of quantum fields can be characterized in a way that
generalizes to curved spacetime. Thus, the spectrum
condition used in Minkowski spacetime can be satisfactorily replaced by a
{\it microlocal spectrum condition} in curved spacetime;
see~\cite{bfk,bf, hw} for further discussion.

However, it is much less obvious how to find a suitable replacement
for property (iii) in curved spacetime. In Minkowski spacetime, the
existence of a unique, Poincare invariant state has very powerful
consequences, so it is clear that a key portion of the content of
quantum field theory in Minkowski spacetime would be missing if we
failed to impose an analogous condition in curved spacetime.  However,
we do not believe that condition (iii) can be generalized to curved
spacetime by a condition that postulates the existence of a preferred
state with special properties.

Very recently, we have proposed~\cite{hw} that the appropriate replacement
of property (iii) for quantum field theory in curved spacetime is to
postulate the existence of a suitable {\em operator product expansion} (OPE)
of the quantum fields. By an OPE, we mean a family of formulae of the form
\ben
\label{OPEidea}
\left\langle\phi^{(i_1)}(x_1) \cdots \phi^{(i_n)}(x_n)\right\rangle_\omega \approx \sum_{j} C^{(i_1)
\dots (i_n)}_{(j)}(x_1, \dots, x_n; y) \, \left\langle \phi^{(j)}(y)
\right\rangle_\omega \, .
\een
Here, the $\phi^{(i)}$ denote the complete collection of fields in the
theory (which may be of arbitrary tensorial or spinorial type),
including all composite fields. The symbol $\langle \,\,\, \rangle_\omega$
denotes the expectation value in the state $\omega$.  Each OPE
coefficient $C^{(i_1) \cdots (i_n)}_{(j)}$ is a distribution on
$M^{n+1}$ that is defined in some open neighborhood of the diagonal in
$M^{n+1}$.  The symbol ``$\approx$'' in eq.~(\ref{OPEidea}) means that
this equation holds in a suitably strong sense as an asymptotic
relation in the limit that $x_1, \cdots, x_n \rightarrow y$. A precise
definition of what is meant by this asymptotic relation is given
in~\cite{hw}.

It was proposed in~\cite{hw} that a quantum field theory should be viewed
as being constructed from the list of quantum fields $\phi^{(i)}$
together with the family of all OPE coefficients
\ben
\label{family}
\C(\M) \equiv \bigg\{ C^{(i_1) \cdots (i_n)}_{(j)}(x_1, \dots, x_n; y) \bigg\} \, ,
\een
where $\M$ denotes all of the background structure, i.e., the
spacetime $(M,g)$ together with choices of space and time orientations
and spin structure. Given $\C(\M)$, the quantum {\em field algebra} of
observables, $\A(\M)$, is then constructed by starting with the free
algebra ${\rm Free}(\M)$ generated by the smeared fields
$\phi^{(i)}(f)$ and factoring it by certain relations. These
relations consist of some ``universal'' relations that do not depend
on the particular theory under consideration (such as linearity of
$\phi^{(i)}(f)$ in $f$) together with certain relations that arise
from the OPE. The precise construction of $\A(\M)$ is given
in~\cite{hw}.  The {\em state space} $\S(\M)$ is then defined to be
the subspace of the space of all linear, functionals $\omega: \A(\M)
\to \mc$ that are positive in the sense that $\omega(A^* A) \equiv
\langle A^* A \rangle_\omega \ge 0$ for all $A \in \A(\M)$, that
satisfy a microlocal spectrum condition, and that satisfy the OPE
relations, eq.~(\ref{OPEidea}).

The collection of OPE coefficients $\C(\M)$ is, of course, not
arbitrary but must satisfy certain general properties, which, in
effect, become the ``axioms'' of quantum field theory in curved
spacetime. The key properties that the OPE coefficients are required
to satisfy include the following: (1) Each $C^{(i_1) \cdots
  (i_n)}_{(j)}(x_1, \dots, x_n; y)$ must be locally and covariantly
constructed from the background structure $\M$. (2) Each $C^{(i_1)
  \cdots (i_n)}_{(j)}(x_1, \dots, x_n; y)$ must satisfy a microlocal
spectrum condition.  (3) The coefficient,
$C^{(i)(i^\star)}_{(\myid)}$,
of the identity element, $\myid$,
must be the most singular OPE
coefficient appearing on the right side of eq.~(\ref{OPEidea}) in the
expansion of $\phi^{(i)}(x_1) \phi^{(i)*}(x_2)$. Furthermore, for $i
\neq \myid$, this coefficient must be singular in the sense of having
positive scaling degree as $x_1, x_2 \rightarrow y$. (4) If we let
$x_1, \cdots, x_n \rightarrow y$ at different rates, then the
$C^{(i_1) \cdots (i_n)}_{(j)}(x_1, \dots, x_n; y)$ must satisfy an
``associativity condition'' corresponding to what one would formally
obtain by first performing an OPE for the subset of points that merge
together the fastest, then performing an OPE on the resulting product
of operators for the (merged) points that merge the second-fastest,
etc.  A precise statement of this condition and a complete enumeration
of all of the other conditions we require for an OPE can be found in~\cite{hw}.

The type of operator product expansion that we require is known to hold
in free field theory and to hold order by order in perturbation theory
for interacting quantum fields in curved
spacetime~\cite{hollands2006}. Thus, what we have proposed can be
viewed as elevating the OPE to the status of a fundamental property of
quantum fields. Although the assumption of the existence of an OPE in
quantum field theory in curved spacetime is remarkably different in
nature from the assumption of the existence of a Poincare invariant
state in quantum field theory in Minkowski spacetime, the OPE plays a role
similar to that of the existence of a Poincare invariant state in
analyses and proofs; roughly speaking the coefficient of the identity,
$C^{(i_1) \cdots (i_n)}_{(\myid)}(x_1, \dots, x_n; y)$, in the OPE in
curved spacetime plays a role similar to that of the vacuum
expectation value $\left\langle 0| \phi^{(i_1)}(x_1) \cdots
\phi^{(i_n)}(x_n)|0 \right\rangle$ in Minkowski spacetime.

Using our new formulation of quantum field theory based on the
existence of a suitable OPE, we have proven~\cite{hw} curved spacetime
versions of the spin-statistics theorem and the PCT
theorem. Interestingly, the PCT theorem in curved spacetime has a
significantly different character than the usual Minkowski
version. The Minkowski version asserts the existence of an
(anti-linear) symmetry associated with the PT isometry of Minkowski
spacetime, which takes quantum fields into their charge
conjugates. The curved spacetime version asserts the existence of a
symmetry relating the quantum field theory defined on an arbitrary
background structure $\M$ to the theory defined on the background
structure $\M'$ obtained from $\M$ by keeping the spacetime manifold,
spacetime metric, and spacetime orientation the same, but reversing
the time orientation. (This symmetry also maps fields to their charge
conjugates.) Thus, for example, the curved spacetime version of the
PCT theorem asserts that for every process that can occur in an
expanding universe, there is a corresponding process (defined by the
PCT symmetry) that occurs in the corresponding {\em contracting}
universe (obtained by reversing the time orientation). We get a ``same
universe'' version of the PCT theorem only in the case of a spacetime
(such as Minkowski spacetime) that admits an isometry that preserves
the spacetime orientation but reverses the time orientation; by
combining the PCT symmetry (which takes $\A(\M)$ to $\A(\M')$) with
the symmetry arising from such a spacetime isometry (which
takes $\A(\M')$ to $\A(\M)$), we obtain a symmetry acting on $\A(\M)$.

However, the potentially most significant ramifications of our new
formulation of quantum field theory in curved spacetime concern the
nature of quantum field theory itself. In our new formulation the
existence of a ``preferred state'' no longer plays any role in the
formulation of quantum field theory. States are inherently non-local
in character, and the replacement of the existence of a preferred
state by the existence of a suitable OPE ---along with the replacement
of Poincare invariance by the condition that the quantum fields be
local and covariant, and the replacement of the spectrum condition by
the microlocal spectrum condition---yields a formulation of quantum
field theory that is entirely local in nature. In this way, the
formulation of quantum field theory becomes much more analogous to the
formulation of classical field theory. Indeed, one can view a
classical field theory as being specified by providing the list of
fields $\phi^{(i)}$ occuring in the theory and the list of local,
partial differential relations satisfied by these fields. Solutions
to the classical field theory are then suitably regular
sections of the appropriate
vector bundles that satisfy the partial differential
relations. Similarly, in our framework, a quantum field theory is
specified by providing the list of fields $\phi^{(i)}$ occuring in the
theory and the list of local, OPE relations satisfied by these fields.
Thus, the OPE relations play a role completely analogous to the role
of field equations in classical field theory. States---which are the
analogs of solutions in classical field theory---are suitably regular
(in the sense of satisfying the microlocal spectrum condition)
positive linear maps on the field algebra that satisfy the OPE
relations. It is worth noting that in classical field
theory, the field equations always manifest all of the symmetries of
the theory, even in cases where there are no solutions that manifest
these symmetries.  Similarly, in our formulation of quantum field
theory, the OPE relations that define the theory should always respect
the symmetries of the theory~\cite{Weinberg}, even if no states happen
to respect these symmetries.

Our viewpoint on quantum field theory is more restrictive than
standard viewpoints in that we require the existence of an OPE. On the
other hand, it is less restrictive in that we do not require the
existence of a ground state. This latter point is best illustrated by
considering a free Klein-Gordon field $\varphi$ in Minkowski
spacetime
\ben
(\Box - m^2) \varphi = 0 \, ,
\een
where the mass term, $m^2$, is allowed to be positive,
zero, or negative. In the standard viewpoint, a quantum field theory
of the free Klein-Gordon field does
not exist in any dimension when $m^2 < 0$ and does not exist in
$D = 2$ when $m^2 = 0$ on account of the non-existence of a Poincare
invariant state. However, there is no difficulty in specifying OPE relations
that satisfy our axioms for all values of $m^2$ and all $D \geq 2$.
In particular, for $D=4$
we can choose the OPE-coefficient $C$ of the identity in the OPE of
$\varphi(x_1) \varphi(x_2)$ to be given by
\bena
\label{kgid}
&&C(x_1, x_2; y) =\\
&&\frac{1}{4\pi^2} \left( \frac{1}{\Delta x^2 + i0t} +
m^2 \, j[m^2 \Delta x^2] \, \log [\mu^2 (\Delta x^2+i0t)] + m^2 \, h[m^2\Delta x^2] \right) \, , \non
\eena
where $\Delta x^2 = (x_1-x_2)^2$ and $t=x^0_1-x^0_2$.
Here $\mu$ is an arbitrarily chosen mass scale and
$j(z) \equiv \frac{1}{2i \sqrt{z}} J_1(i\sqrt{z})$ is an analytic function
of $z$, where $J_1$ denotes the Bessel function of order $1$.
Furthermore, $h(z)$ is the analytic function defined by
\ben
h(z) = -\pi \sum_{k=0}^\infty[\psi(k+1) + \psi(k+2)]\frac{(z/4)^k}{k!(k+1)!} \, .
\een
with $\psi$ the psi-function.
This formula for the OPE coefficient---as well as the corresponding
formulas for all of the other OPE coefficients---is as well defined for
negative $m^2$ as for positive $m^2$. Existence of states satisfying all
of the OPE relations for negative $m^2$ can be proven by the deformation
argument of~\cite{FNW}, using the fact that such states
exist for positive $m^2$.

The potential
importance of the above example is that it explicitly demonstrates
that the local OPE coefficients can have a much more regular behavior
under variations of the parameters of the theory as compared with
state-dependent quantities, such as vacuum expectation values. The OPE
coefficients in the above example are analytic in $m^2$.  On the other
hand, the 2-point function of the global vacuum state is, of course,
defined only
for $m^2 \geq 0$ and is given by
\bena
\label{kgvev}
&&\langle 0| \varphi(x_1) \varphi(x_2)|0 \rangle  =\\
&&\frac{1}{4\pi^2} \left( \frac{1}{\Delta x^2 +i0t} +
m^2 \, j[m^2 \Delta x^2] \, \log [m^2 (\Delta x^2+i0t)]
+ m^2 \, h[m^2\Delta x^2]
\right) \, . \non
\eena
This behaves non-analytically in $m^2$ at $m^2 = 0$ on account of the
$\log m^2$ term. In other words, in free Klein-Gordon theory, vacuum
expectation values cannot be constructed perturbatively by expanding
about $m^2 = 0$---as should be expected, since no vacuum state exists
for $m^2 < 0$---but there is no difficulty in perturbatively constructing
the OPE coefficients by expanding about $m^2 = 0$.

In quantum field theory in Minkowski spacetime, attention is usually
focused upon quantities that involve states, such as S-matrix
elements. Since the states of interest cannot be expected to vary
analytically with the parameters of the theory, perturbation
expressions for quantities such as S-matrix elements cannot be
expected to converge, and there is ample evidence that they do
not. However, the above considerations suggest the possibility that
the OPE coefficients may have much better behavior than quantities
associated with states, and that the perturbation series for the OPE
coefficients may converge. In other words, we are suggesting the
possibility that within our framework, it may be possible to
perturbatively construct interacting quantum field
theories\footnote{In order to do so, it will be necessary to define
  the basis fields $\phi^{(i)}$ appropriately and also to parametrize
  the theory appropriately (since a theory with an analytic dependence
  on a parameter could always be made to appear non-analytic by a
  non-analytic reparametrization).}.  Aside from the free Klein-Gordon
example above, the only evidence we have in favor of convergence of
perturbative expansions for OPE coefficients is the example of
super-renormalizable theories, such as $\lambda \varphi^4$-theory in
two spacetime dimensions~\cite{hollandskopper}. Here, only finitely
many terms in a perturbative expansion can contribute to any OPE
coefficient up to any given scaling degree, so convergence (up to any
given scaling degree) is trivial. By contrast, for $\lambda
\varphi^4$-theory in two spacetime dimensions, the rigorously
constructed, non-perturbative ground state $n$-point functions can be
proven to be non-analytic at $\lambda=0$; see e.g.~\cite{Rivasseau}. Of course,
even if we had the complete list of OPE coefficients, we would still
need to construct states, which cannot be done
perturbatively. Nevertheless, it would be potentially very useful to
have the OPE coefficients even if one did not have states (or even an
existence proof for states)---just as in classical field theory it is
useful to have the field equations even if one does not have a method
for finding solutions.

In cases---such as free Klein-Gordon theory above---where the OPE
coefficients {\it can} be chosen to be analytic in the parameters of
the theory, it seems natural to {\it require} that the theory be
defined so that this analytic dependence holds. This requirement has
some potentially major ramifications, which we now discuss.
Since a vacuum expectation value of
a products of fields (i.e., a
correlation function) would be expected to have a
non-analytic dependence on the parameters of the theory, it follows that
if the OPE coefficients have an analytic dependence on these
parameters, then, even in Minkowski spacetime,
some of the fields appearing on the right side
of the OPE eq.~(\ref{OPEidea})
must acquire a nonvanishing vacuum expectation value, at least for
some values of the parameters. This point is well illustrated by the above
Klein-Gordon example. It is natural to identify the next term
(i.e., the
term beyond the identity term)
in the OPE of $\varphi(x_1) \varphi(x_2)$
as being $\varphi^2$ (with unit coefficient), i.e.,
\ben
\varphi(x_1) \varphi(x_2) \sim C(x_1, x_2; y) \myid
+ \varphi^2 (y) + ... \,\, ,
\label{ps}
\een
with $C(x_1, x_2; y)$ given by eq.~(\ref{kgid}).
This corresponds to the
usual ``point-splitting'' definition of $\varphi^2$, except that
$C(x_1, x_2; y)$ now replaces
$\langle 0|\varphi(x_1) \varphi(x_2) |0 \rangle$.
If we take
the vacuum expectation value of this formula (for $m^2 \geq 0$, when a
vacuum state exists)
and compare it with
eq.~\eqref{kgvev}, we obtain
\ben
\langle 0| \varphi^2 |0 \rangle = -\frac{m^2}{16\pi^2} \log (m^2/\mu^2) \, .
\een
Thus, we cannot set $\langle 0| \varphi^2 |0 \rangle = 0$ for all values
of $m^2$.
A similar calculation for the stress-energy tensor of $\varphi$ yields
\ben
\langle 0| T_{ab} |0 \rangle = \frac{m^4}{64\pi^2} \log (m^2/\mu^2) \eta_{ab} \, .
\label{Tab}
\een

As in other approaches, the freedom to choose the arbitrary
mass scale $\mu$ in eq.~\eqref{kgid} gives rise to a freedom to
choose the value of the ``cosmological constant term''
in eq.~\eqref{Tab}.  However, unlike other
approaches, there is no freedom to adjust the value of the
cosmological constant when $m^2 = 0$ (i.e., we unambiguously obtain
$\langle 0| T_{ab} |0 \rangle = 0$ in Minkowski spacetime
in that case), and the $m^2$-dependence
of the cosmological constant is fixed (since $\mu$ is not allowed to
depend upon $m^2$).

Although the example of the free Klein-Gordon field is, of course, too
trivial to be realistic, it serves to illustrate the conflict between
the expected non-analytic behavior of the left side of
eq.~(\ref{OPEidea}) and the conjectured analytic behavior of the OPE
coefficients---a conflict that can be resolved only if the operators
appearing on the right side of eq.~(\ref{OPEidea}) generically acquire
a nonvanishing vacuum expectation value.  A much more interesting
example arises for interacting field theories, such as non-abelian
gauge theories, where ``nonperturbative'' effects are known to
arise. If such nonperturbative effects contribute at finite scaling
degree to the field
correlation functions appearing on the left side of
eq.~(\ref{OPEidea}), then it is natural to expect that they will
similarly contribute to the vacuum expectation values of the fields
appearing on the right side. In particular, they may contribute to
the vacuum expectation value of the stress-energy tensor.

One of the great mysteries of modern cosmology is to account for the
acceleration of the present universe. In order to explain the observed
acceleration, one must postulate the existence of ``dark energy'', a
component of matter that is distributed uniformly throughout the
universe and has large negative pressure. There is growing evidence
that ``dark energy'' corresponds to a cosmological constant term in
Einstein's equation, i.e., a stress-energy tensor proportional to the
metric. While it is not difficult to imagine how a ``vacuum energy''
contribution of this general form to the stress-energy tensor could
arise, it is very difficult to imagine how one could account for the
incredible mis-match of scales between the value of the cosmological
constant required to explain the observed acceleration---corresponding
to a length scale of order the Hubble radius---and the natural length
scales occurring in particle physics. We are proposing
that this mis-match might be explained if, as we have
argued above, the ``vacuum energy'' is associated with
non-perturbative effects, since non-perturbative effects can
potentially be extremely small compared with the natural scales
appearing in a theory. This possibility appears worthy
of further investigation.

\bigskip

\noindent
{\bf Acknowledgments:}
This research was supported in part by NSF Grant PHY04-56619 to the University
of Chicago.

\end{document}